# Combined Quark-Lepton-Complementarity via United by One-Parameter Particle Mixing Angles

E. M. Lipmanov
40 Wallingford Road # 272, Brighton MA 02135, USA

**Abstract**

On the level of phenomenology unification-type idea of a basic connection between dimensionless mass and charge quantities requires particle mass copies and therefore is interesting as raison d'etre for particle flavor (degree of freedom). Particle mass ratios and mixing angles expressed in terms of related to the fine structure constant $\varepsilon$-parameter are particular examples of mass-matrix -- charge unification. In this paper small violations of quark-lepton-complementarity (QLC) come from an accurate system of three quark mixing angles in terms of the $\varepsilon$-parameter. Two patterns of small QLC-violations, 'combined' and 'universal', are considered. The first pattern predicts solar mixing angle $\theta_{sol} \cong 34.04°$ as ~7% violation of QLC, and atmospheric mixing angle $\theta_{atm} \cong 42.64°$ that is in agreement with QLC. The 'universal' pattern predicts the same magnitude of solar angle, while the atmospheric one violates QLC by ~1%, $\theta_{atm} \cong 43.0°$. Quantitative tests of both QLC-violation patterns and choice between them can be made at accurate neutrino oscillation experiments.

_________

**1.** Quark-lepton complementarity (QLC) is an empirically indicated geometric rule approximately connecting the two largest quark CKM and neutrino PMNS mixing angles [1] that is widely discussed in the literature.

In the framework of benchmark flavor pattern [2], a perturbation by the new small universal flavor-electroweak $\varepsilon$-parameter [3] naturally leads to exact geometric QLC at leading $\varepsilon$-approximation. In point of fact, it is a connection between the two largest quark and lepton mixing angles via universal particle charge since by definition the $\varepsilon$-parameter,

$$\varepsilon = \exp(-5/2), \qquad (1)$$

is related to electric charge absolute value

$$e \cong \varepsilon \sqrt{(hc)}. \qquad (2)$$

That connection has a basic physical meaning. At considered level of phenomenology the idea of a connection between universal electric charge and lepton and quark dimensionless mass-matrix quantities means raison d'etre for particle flavor (degree of freedom) because the realization of that idea [4] is possible only if there are particle mass copies.

**2.** At leading $\varepsilon$-approximation [2] the quark mixing angles are

$$\sin^2(2\theta_c) = 2\varepsilon, \quad \theta_c \cong 11.9°, \qquad (3)$$

$$\sin^2(2\theta_{23}) = \varepsilon^2, \quad \theta_{23} \cong 2.4°, \qquad (4)$$

$$\sin^2(2\theta_{13}) = \varepsilon^4, \quad \theta_{13} \cong 0.19°. \qquad (5)$$

An accurate extension of this pattern to next to leading $\varepsilon$-approximation by exponential factors [4] is given by

$$\sin^2(2\theta_c) \cong f(2\varepsilon) = (2\varepsilon)\exp(2\varepsilon), \quad \theta_c \cong 13.047°, \qquad (6)$$

$$\sin^2(2\theta_{23}) \cong f(\varepsilon^2) = (\varepsilon^2)\exp(\varepsilon^2), \quad \theta_{23} \cong 2.362°, \qquad (7)$$

$$\sin^2(2\theta_{13}) \cong f(\varepsilon^4) = (\varepsilon^4)\exp(\varepsilon^4), \quad \theta_{13} \cong 0.193°. \qquad (8)$$

A simple universal function is introduced here,

$$f(x) = x \exp(x), \tag{9}$$

it emphasizes the regularity of the system (6)-(8). Predictions (6)-(8) are accurate to within the small 1 S.D. of the quark CKM world fit mixing matrix [5].

The system of three quark mixing angles (6)-(8) is well suited for describing relations between quark and lepton mixing angles that respect approximate QLC in a quite definite way.

**3.** 'Combined' QLC-violation pattern differs from the exact QLC one by change sign of the $\varepsilon$-parameter

$$\sin(\ldots) \to \cos(\ldots), \quad \varepsilon \to (-\varepsilon). \tag{10}$$

When applied to the two largest quark mixing angles (6) and (7), definition (10) determines the $\varepsilon$-structures and magnitudes of neutrino oscillation solar and atmospheric mixing angles

$$\cos^2(2\theta_{sol}) = |f(-2\varepsilon)|, \quad \theta_{sol} \cong 34.04°, \tag{11}$$

$$\cos^2(2\theta_{atm}) = f(\varepsilon^2), \quad \theta_{atm} \cong 42.64°. \tag{12}$$

The relations (11)-(12) present a unique form of small QLC-violation that is fully expressed through the universal function (9) in analogy with quark mixing angle relations (6)-(7).

Compare (11)-(12) with taken for granted best-fit values of neutrino mixing angles e.g. [6],

$$(\sin^2 2\theta_{sol})_{exp} = 0.312 + 0.063 - 0.049, \quad (\theta_{sol})_{bf} = 33.96°, \tag{13}$$

$$(\sin^2 2\theta_{atm})_{exp} = 0.466 + 0.178 - 0.135, \quad (\theta_{atm})_{bf} = 43.05°. \tag{14}$$

As a result, the solar and atmospheric angles (11) and (12) fit the data values by ~0.2% and ~1% respectively.

If QLC were an exact regularity, the solar and atmospheric neutrino angles would be $\theta_{sol} \cong 31.95°$ and $\theta_{atm} \cong 42.64°$. So, the atmospheric neutrino mixing angle (12) obeys QLC while the deviations from exact QLC of the solar angle is large ~7%.

**4.** 'Universal' QLC-violation as discussed in [4] is defined by relations:

$$\cos^2(2\theta_{sol}) = f(2\varepsilon)\,\varepsilon^{-4\varepsilon}, \quad \theta_{sol} \cong 34.04^o, \tag{15}$$

$$\cos^2(2\theta_{atm}) = f(\varepsilon^2)\,\varepsilon^{-4\varepsilon}, \quad \theta_{atm} \cong 43.0^o. \tag{16}$$

It leads to the same result (11) for solar mixing angle, but it may be in slightly better agreement with the experimental indications for atmospheric mixing angle. As a remark, in contrast to combined QLC-violation pattern (11)-(12) the relation (16) for atmospheric angle cannot be fully expressed through the universal function (9).

Notice that in both considered QLC-violation patterns (11)-(12) and (15)-(16) the limit $\varepsilon \to 0$ means exact QLC relations in accordance with the definition of benchmark flavor pattern [2].

Both quantitative QLC-extensions (11)-(12) and (15)-(16) and the difference between the two of them can be decisively tested at coming accurate neutrino oscillation experiments.